\documentstyle[prb,aps,psfig]{revtex}

\newcommand{\half}{\frac 1 2 }
\newcommand{\eg}{{\em e.g.} }
\newcommand{\ie}{{\em i.e.} }
\newcommand{\etc} {{\em etc.}}

\newcommand{\noi}{\noindent}
\newcommand{\etal}{{\em et al.\ }}

\newcommand{\pref}[1]{(\ref{#1})}

\newcommand {\be}[1]{
      \begin{eqnarray} \mbox{$\label{#1}$}}

\newcommand{\ee}{\end{eqnarray}}

\def\phidag{\phi^{\dagger}}

\def\chidag{\chi^{\dagger}}

\def\aeo{a_1^0}
\def\ato{a_2^0}
\def\aev{\vec a_1}
\def\atv{\vec a_2}
\def\sz{\, \vec\sigma\cdot \hat m}

\def\at{\tilde a}
\def\atvidv{\vec{\tilde a}}
\newcommand{\ki}[2]{l_{#1 #2}}

\def\rhobar{\bar\rho}

\def\mh{\hat m}
\def\rup{\rho_{\uparrow}}
\def\rdown{\rho_{\downarrow}}
\def\p{\partial}
\def\zb{\bar z}
\def\Pbar{\bar P}

\begin{document}
\centerline{\Large\bf Edge theories for polarized quantum Hall states}
\vspace* {-21 mm}
\begin{flushright}
  NORDITA-1999/41 CM \\
  USITP-99-5 \\
  July-1999 \\
\end{flushright}

\vskip 20 mm
\centerline{\bf  T.H. Hansson$^*$  and S. Viefers$^{\dagger}$ }
\vskip 8 mm
\begin{center}
{\em $^*$ Department of Physics, University of Stockholm, Box 6730,
S-11385 Stockholm, Sweden } \\ 
\vskip 3mm
{\em $^{\dagger}$ NORDITA, Blegdamsvej 17, DK-2100 Copenhagen {\O}, Denmark} \\
\end{center}

\vskip 10mm

\centerline{\bf ABSTRACT}
\vskip 3mm
Starting from recently proposed bosonic mean field theories for
fully and partially polarized
quantum Hall states, we construct  corresponding effective low
energy theories
for the edge modes. The requirements
of gauge symmetry and invariance under global O(3) spin rotations, broken
only by a Zeeman coupling, imply boundary conditions that allow for
edge spin waves. In the generic case, these modes are chiral, and the
spin stiffness differs from that in the bulk. For the case of
a fully polarized $\nu=1$ state, our
results agree with previous Hartree-Fock calculations.

\vskip 15mm

{\begin{center}
{\bf 1. INTRODUCTION} 
\end{center}
\renewcommand{\theequation}{1.\arabic{equation}}
\setcounter{equation}{0}
\vskip 5mm
The study of edge modes in the quantum Hall (QH)
system is motivated by their importance  for the charge transport,
and also because they are believed to give important
information about the bulk state.

A class of models which have been quite successful in describing the bulk
QH states, are the Chern-Simons Ginzburg-Landau
(CSGL) effective field theories.  The simplest form of such a theory
 describes  spinless QH systems at Laughlin filling fractions
$\nu = 1/m$ \cite{zhang1}, but generalizations can also
describe more complicated filling fractions, multilayer systems
and effects of spin \cite{wen1,lee1}.
Starting from a CSGL theory, it is possible to construct effective,
one-dimensional chiral boson models describing the low energy physics
at the edge.
These models, both for the simple single-component spinless case and
more general multicomponent cases, have been analyzed by Wen
\cite{wen1} and others.

The QH states are incompressible, \ie there is a finite energy gap to
density fluctuations in the bulk.
Because of the magnetic field, the ground state
is usually at least partially polarized, and the spectrum also
includes ferrromagnetic spin waves. Generically, one would expect the
Zeeman gap to be very large, but material effects can renormalize the
$g$-factor to small enough values for the gap to be comparable
to typical Coulomb energies. Thus it is
interesting to study polarization effects, both in bulk and at the edge.

The perhaps most striking  polarization effect is that, under certain
conditions, the lowest charged excitation in a $\nu = 1$ state is a
topological soliton, a so-called skyrmion, as originally predicted by
Sondhi \etal \cite{sondhi1}.

Concerning the ground states, there is strong experimental
evidence that at low electron densities fractional QH
ground states at certain fractions such as $2/5$ and $4/3$ are not fully
polarized. This has been established by studying the transport properties at
a fixed filling fraction as a function of the applied magnetic field $\vec B$
(either by tilting the field or by changing the electron
density)\cite{clark1,eisenstein1,syphers1}.

There have also been many
studies of spin polarization effects at the edge of QH-states, and in
particular Karlhede \etal, using Hartree-Fock techniques and
an effective sigma-model, predicted that for smooth enough potential the
edge would reconstruct to a spin-textured state \cite{karlhede1}.
Similar results were obtained by Oaknin \etal for quantum
dots\cite{oaknin1} and by  Leinaas and Viefers \cite{leinaas1}, who solved a
full CSGL model of fully polarized QH states \cite{lee1}
in the presence of an infinite edge.

Several papers have  studied the interplay between density and
spin fluctuations at the edge and derived the pertinent dispersion
relations \cite{leinaas1,karlhede2,franco1}. In a very recent
TDHF-study of the $\nu = 1$ state Karlhede, Lejnell and Sondhi \cite{karlhede3}
find that in
addition to the edge magnetoplasmons and bulk spin waves, there are
also edge spin waves, which generically are chiral.

With this background, it is clearly interesting to construct a
class of  effective theories that describe the low
energy physics both in the bulk and at the edge, and which are general
enough to describe polarization effects around both fully and
partially polarized ground states.

This is the aim of the present paper, where we start from
a recently proposed effective CSGL model  describing fully and
partially polarized states \cite{hansson1}.
In addition to a two-component bosonic
field $\phi$ and two Chern-Simons (CS) gauge fields,
which together describe the electrons, this
model also contains a $\sigma$-model field, $\hat m$, describing the spin
polarization direction. In the special case of full polarization,
the model only involves a single scalar field $\phi$ and one CS field.

The main result of our analysis is an effective action consisting of
a bulk part that is derived from the CSGL model by integrating out
the high energy modes, and an edge part which is essentially a chiral
boson Lagrangian of the type originally proposed by Wen \cite{wen1}, but
coupled to the bulk in such a way that not only the total charge
current, but also the total spin current is conserved (up to an
explicit breaking due to the Zeeman effect). We analyze the small
fluctuation spectrum of this model and find, in addition to the
expected edge magnetoplasmons, also (gapped) edge spin waves  which
generically are chiral and, for soft edge potentials,  propagate in
the same direction as the edge magnetoplasmons.

The paper is organized as follows. In the next section we give
the effective bulk Lagrangians together with a short physical motivation
for their general form. The actual derivation from CSGL theories, which is
rather technical, is referred to Appendix A.
In the following two sections we construct the
corresponding edge theories and show how the presence of edge currents
and charge densities implies a boundary condition on the sigma model
field describing the spin.  Section 5 contains an analysis  of the
excitation spectrum of our model with emphasis on the edge spin waves.
Here we also compare with the work of Karlhede \etal\cite{karlhede3} and
Milovanovi\'c \cite{milo1}. We summarize our results in section 6,
and an alternative way to derive the effective
low-energy model for partially polarized states, using a dual
CSGL formulation, is given in  Appendix B.

\vskip 4mm
\begin{center}
{\bf 2. CHERN-SIMONS BULK LAGRANGIANS }
\end{center}
\renewcommand{\theequation}{2.\arabic{equation}}
\setcounter{equation}{0}
\vskip2mm

\newcommand{\ab}{\bar\alpha}
\newcommand{\ca}{\cos\alpha}
\newcommand{\sa}{\sin\alpha}
\newcommand{\tho}{\vartheta}
\newcommand{\tht}{\theta}
\newcommand{\eone}{\hat e_1}
\newcommand{\etwoz}{\hat e_2^0}
\newcommand{\eonez}{\hat e_1^0}
\newcommand{\etwo}{\hat e_2}
\newcommand{\mhz} {{\hat m^0}}

Consider a two-dimensional electron gas subject to a magnetic field of
constant magnitude and direction, $\vec B = \nabla \times \vec A_{bg}$.
In addition we probe the system with a weak external electromagnetic
potential
$\delta A^\mu$, so the total potential is
$A^i = A^i_{bg} + \delta A^i$ and
$A^0 = \delta A^0$. In the following we shall
include the electron charge in the gauge potentials, \ie to get
back to physical
units one should make the replacement $A_\mu \rightarrow -eA_\mu$,
where $e>0$.
In particular this means that electron densities and charge densities
are both
positive, and since the physical background magnetic field is taken in the
positive $z$-direction,  the background field
$B_{\perp} \equiv \epsilon^{ij}\p_i A_{j,bg}$ will be negative.
We also put $\hbar =c =1$. Here and in the following repeated indices
are summed over, but there is no distinction between upper and lower
indices, and all signs are written explicitly (in particular, note
that the  potential energy is   $-A^0$).

At the special filling fractions corresponding to QH plateaux, the electrons
form incompressible quantum liquids, characterized by quantized Hall
conductivities, $\sigma_H$. The electromagnetic response of such states is
described by effective actions of the CS type; in the simplest case
(corresponding to spinless electrons in the principal Laughlin states)
the effective low-energy Lagrangian reads
\be{ncsact}
{\cal L}^{sl}_{eff} = - \frac l {4\pi}\epsilon^{\mu\nu\sigma}
                        A_{\mu} \partial_{\nu} A_{\sigma}
                    \equiv - \frac l {4\pi} AdA  \ \ \ \ \  ,
\ee
where $l^{-1}$ is an odd integer. The corresponding current is given by
\be{nbcurr1}
J_\mu \equiv \frac {\delta S} {\delta A^\mu}
          = - \frac l {2\pi} \epsilon_{\mu\nu\sigma}\partial^{\nu} A^{\sigma}
                  \ \ \ ,                                  
\ee
with the correct Hall conductivity, $\sigma_H = l/2\pi$.

We shall now consider systems where the Zeeman gap is
small so that spin waves should be included in the effective
low energy theories;
the Lagrangians will thus depend on a dynamical $\sigma$-model field $\mh$,
describing the spin polarization direction,  in addition
to the electromagnetic field $A$. In this section we give some simple
arguments to motivate the form of the Lagrangians and then simply state the
final results.  The actual derivation  from CSGL-theories is outlined in
Appendix A.

Since we consider systems in a strong magnetic field, we expect the
elecrons to be (at least partially) polarized. Thus, the ground state is
ferromagnetic, and the dynamics of the spin field $\mh$ is described by
an O(3) $\sigma$-model. It would, however, be too simplistic just to add
a  $\sigma$-model Lagrangian to \pref{ncsact}, since there is a non-trivial
connection between spin and charge due to the strong magnetic field.
This can be understood in many ways \cite{sondhi1,hansson1,moon1}, \eg by calculating
the Berry phase corresponding to adiabatic motion of a single electron in
a (static) polarized background of the other electrons.
Assuming that the spin of the electron remains aligned with that of the
background at all times, the Berry phase can be expressed as an
Aharonov-Bohm phase due to a {\em topological} vector potential $\tilde a$,
related to the spin mean field $\mh$ via
\be{natilde}
\tilde a^\mu = \chidag_{\mh} i\partial^\mu \chi_{\mh}
 \ \ \ \ \ ,
\ee
where the spinor $\chi_{\mh}$ is  defined by
\be{nchi}
\hat m^i = \chidag_{\mh} \sigma^i \chi_{\mh} \ \ \ \ \ ,
\ee
up to a gauge transformation. For fully polarized (fp) states,
the electron cannot distinguish between
the real electromagnetic potential $A$ and the topological potential
$\tilde a$, so they can appear only in the combination $A+\tilde a$.
In this case
the effective Lagrangian is given by
\be{neff1a}
{\cal L}^{fp}_{eff} = -\frac l {4\pi}  (A + \tilde a )d(A + \tilde
a) - \frac {\kappa\rhobar} 4 (\vec\nabla\mh)^2  \ \ \ \ \  ,
\ee
with the corresponding current
\be{fpc}
J_{\mu} =  - \frac l {2\pi} \epsilon_{\mu\nu\sigma}\partial^{\nu}
            (A^{\sigma} + {\tilde a}^{\sigma}) \ \ \ \ \ ,
\ee
where we denoted the ground state density by
$\rhobar$ , \ie $\rhobar = \langle
J_{0}\rangle$. (In this context, we can replace $J_{0}$ by $\rhobar$ since the
difference is given by higher derivative terms, as discussed in Appendix A.)
It is not obvious that the Lagrangian \pref{neff1a} describes a
$\sigma$-model, but in
fact it does.
The CS-part  contains a term $\sim B_\perp
\tilde a^0$ which is nothing but the kinetic part of the $\sigma$-model, and
the Zeeman term is included as a spin-dependent part of the
scalar potential \ie
$A^0\rightarrow A^{0} - (\mu_{e}/2)\vec B\cdot\mh$.
For further details we again refer to Appendix A.
The Lagrangian \pref{neff1a} has also been discussed by Baez 
{\em et al.}\cite{baez1}.

Partially polarized (pp) ground states are characterized by two conserved
charges, being the total number of electrons with spin
pointing along and opposite
to the mean field, $\mh$,  respectively.
Assuming that  these charges are  separately conserved  amounts to
neglecting the effects of spin-flip transitions.
A spin wave corresponds to fluctuations
in the direction of $\mh$, while keeping the local polarization
$P$ fixed at its mean field value, $\Pbar$ (see eq.\pref{gnst}).
Corresponding to the two conserved charges,
there are now two gauge symmetries:
The first is related to electromagnetism, described
by the potential $A \equiv A_1$, and the second is related to the Berry
phase, \ie the potential $\tilde a \equiv A_2$.
Here, we have introduced the convenient notation $A_\alpha$,
$\alpha =1,\, 2$ for the two gauge fields.
In this case the effective Lagrangian reads
\be{neff2}
{\cal L}^{pp}_{eff} =  -\frac 1 {2\pi} \ki \alpha\beta A_\alpha d A_\beta
- \frac {\kappa\rhobar\Pbar} 4 (\vec\nabla\mh)^2
 \ \ \ \ \ ,
\ee
where the form of the matrix $\bf l$ is given in Appendix A. Again,
the kinetic term for the $\sigma$-model is hidden in the CS-term,
and the Zeeman term is included by a shift in the potential $A_2^0$.
The two conserved currents corresponding to \pref{neff2} are
\be{nbcurr2}
J_\alpha^\mu  = - \frac 1 \pi \ki\alpha\beta
              \epsilon^{\mu\nu\sigma} \p_{\nu}
              A_{\sigma}^\beta \ \ \ \ \  .
\ee
In addition to the gauge symmetries of the fully- and partially polarized
models, there is, in the absence of a Zeeman interaction, a global O(3)
symmetry corresponding to rotations of the spin field $\mh$. In the
following we shall assume that, even in the presence of an edge, the only
explicit breaking of this symmetry is due to the Zeeman term.

The effective low-energy Lagrangians \pref{neff1a} and \pref{neff2}
are the starting point for the next section, where we construct the
corresponding edge Lagrangians.

\vskip 4mm
\begin{center}
{\bf 3. EDGE THEORIES }
\end{center}
\renewcommand{\theequation}{3.\arabic{equation}}
\setcounter{equation}{0}
\vskip2mm

In principle one should be able to derive the edge theory directly
from the microscopic physics, and for the case of a spin polarized
$\nu = 1$ state this was done by Stone\cite{stone2}. For general
filling fractions and polarizations, a possible route would be to
start from a CSGL formulation and use mean field theory to
find an explicit  ground state solution in
the presence of an edge potential.
Then one would try to extract an effective edge Lagrangian.
A step in this direction was taken by
Leinaas and Viefers \cite{leinaas1} who (numerically) found
solutions to the CSGL mean field equations,
showing the existence of edge spin
textures, and further studied the gapless edge excitations
(edge magnetoplasmons) of the system.
In this paper we shall not take this approach, but
instead, following Wen and others \cite{wen1,baez1}, 
construct the edge theories
using general arguments based on symmetry, incompressibility and topological
order. In the spinless case, these properties
imply the existence of massless chiral degrees of
freedom at the edge of the quantum Hall droplet.
These edge modes are described by chiral boson theories, where
the charges reflect the topological order in the bulk, and the velocities
of the different modes are non-universal parameters depending on
the edge potentials.

An important aspect of the edge theories is that they are needed to cancel the
gauge non-invariant terms that appear in CS-theories in the presence of
boundaries.
(With a common abuse of terminology we shall refer to these terms
as ``anomalies''.)
To be specific, we shall use a straight edge geometry,
where the bulk is in the upper half-plane, and the edge is at $y = 0$.
This is made explicit by writing the bulk action as
$S = \int d^3 r~\theta(y){\cal L}_{eff}$, where $\theta(y)$ is the
step function and ${\cal L}_{eff} $ is given by \pref{neff1a} or \pref{neff2}.
Under the gauge-transformation
\be{gauge}
A_\alpha \rightarrow A_\alpha - d\epsilon_\alpha \ \ \ \ \ ,
\ee
we get the anomalies
\be{an}
\delta S^{fp} &=& \frac l {4\pi}
     \int d^3 r~ \delta(y) (E^{x} + \tilde e^{x}) \epsilon_{1}
\label{an1} \\
\delta S^{pp} &=& \frac 1 {2\pi}l_{\alpha\beta}
     \int d^3 r~ \delta(y) E_{\alpha}^{x} \epsilon_\beta
\label{an2} \ \ \ \ \ ,
\ee
where $\vec E$, $\vec{\tilde e}$ and $\vec  E_\alpha$ are the electric fields
corresponding to the potentials $A$, $\tilde a$ and $A_\alpha$ respectively,
$E_x = -\epsilon_{ab}\p_a A_b$ (the indices $a,b$ denote the edge
coordinates $(0,x)$).

The chiral bosons on the edge are described by real scalar phase fields
$\phi$,
which couple to external gauge fields in a gauge invariant combination provided
the scalars transform as
\be{phi}
\phi \rightarrow \phi + \epsilon \ \ \ \ \ .
\ee
For the fully polarized case, the corresponding covariant
derivatives are thus given by
\be{covdir}
\Delta^0 \phi &=& \partial^0 \phi + A^{0} + \tilde a^0 \\
\Delta^x \phi &=& \partial^x \phi + A^x + \tilde a^x  \ \ \ \ \ . \nonumber
\ee
For the spinless case we just omit the $\tilde a_{\mu}$ term.
For partially polarized states, two chiral bosons are needed in order
to cancel the two anomalies \pref{an2}, and we have
\be{ppcovdir}
\Delta^0 \phi_\alpha &=&  \partial^{0} \phi_\alpha + A^0_\alpha \\
\Delta^x \phi_\alpha &=& \partial^{x} \phi_\alpha + A^x_\alpha
\ \ \ \ \ . \nonumber
\ee
The edge Lagrangians are now constructed as a sum over a
gauge-invariant part that,
for zero external potentials, reduces
to the edge theories given by Wen \cite{wen1} and Stone \cite{stone2},
and a gauge-noninvariant part
that cancels the anomalies \pref{an1} and \pref{an2}. Thus we write the total
action as
\be{stot}
S^{tot} = \int d^{3}r\, \left[ \theta (y){\cal L}_{eff} + \delta(y)
{\cal L}_{ed} \right] \ \ \ \ \ ,
\ee
where ${\cal L}_{eff} $ is given by \pref{neff1a} or \pref{neff2}.
It is easy to see that with the choice
\be{ed}
{\cal L}_{ed}^{fp} &=& {\cal L}^{gi}_{ed}(\Delta^{a}\phi)
- \frac l {4\pi} \phi (E_x + \tilde e_x)  \\
{\cal L}_{ed}^{pp} &=& {\cal L}^{gi}_{ed}(\Delta^{a}\phi_{\alpha})
- \frac 1 {2\pi} l_{\alpha\beta} \phi_\alpha E^x_\beta \ \ \ \ \ ,
\nonumber
\ee
where ${\cal L}^{gi}_{ed}$ denotes the gauge invariant part,
$S^{tot}$ is gauge invariant. The {\em fully} polarized case has been
considered in earlier papers\cite{milo1,baez1}. 
Our result \pref{ed} agrees with the work
of Baez {\em at al.} \cite{baez1}, but differs from that of
Milovanovi\'c \cite{milo1}.

The conserved currents associated with
the gauge symmetries have both a bulk and an edge contribution, \ie,
\be{jtot}
J^{tot}_{\mu}
   &=& \theta(y) J_{\mu} + \delta(y) j^{a}
      \delta_{\mu a}    \ \ \ \ \ \ (fp)  \\
J^{tot,\alpha}_{\mu}
   &=& \theta(y) J_{\mu}^{\alpha} + \delta(y) j^{a}_{\alpha}
      \delta_{\mu a} \ \ \ \ \ \ (pp) \ \ \ , \nonumber
\ee
where the bulk currents are given by \pref{fpc} and \pref{nbcurr2},
respectively, and the
edge currents by
\be{j}
j^{a} &=& \frac { \partial {\cal L}^{gi}_{ed} } {\partial A^{a} }
    + \frac l {4\pi} \epsilon^{ab}\Delta_{b}\phi
        \ \ \ \ \ \ \ \ \ (fp) \label{j1} \\
j^{a}_{\alpha} &=& \frac {\partial {\cal L}^{gi}_{ed} } {\partial
A^{a}_{\alpha} } +  \frac 1 {2\pi} l_{\alpha\beta}
   \epsilon^{ab}\Delta_{b}\phi_{\beta}
   \ \ \ \ \ (pp) \ \ \  . \label{j2}
\ee
Some care is needed in deriving these expressions from the action
\pref{stot}. There are contributions both from the bulk and
the edge actions, and only the sum of these is gauge invariant.
Note that although  the total currents \pref{jtot} are conserved,
the edge parts \pref{j1} and \pref{j2} are not, but satisfy
\be{ann}
\partial_{a}j^{a}  &=&  - \frac l {2\pi} (E^{1} + \tilde e^{1} )
                    = - J^{y} \ \ \ \ \ \ \ (fp)  \label{ann1} \\
\partial_{a}j^{a}_{\alpha} &=&  - \frac 1 {\pi}
                   l_{\alpha\beta}E^{1}_{\beta}
       = - J^{y}_{\alpha}  \ \ \ \ \ \ \ \ \ \ \
 (pp)\ \ \  ,  \nonumber
\ee
where the last identities follow from the explicit expressions
\pref{fpc} and \pref{nbcurr2} for the bulk currents.

At this point it is illuminating to recall the well
understood relation between anomaly cancellation and charge
conservation. As stressed by Stone \cite{stone2}, the current
flow from the bulk
into the edge means that the edge
currents are not conserved. This is
easily understood by considering  the region
$[x-\epsilon , x+\epsilon]$
of the edge, where charge conservation is expressed as
\be{globcon}
 {\partial}_{0} \int^{x+\epsilon}_{x-\epsilon}dx'
j^{0}(x') = - j^{x}(x+\epsilon) + j^{x}(x-\epsilon)
          - \int^{x+\epsilon}_{x-\epsilon}dx'  J^{y}(x') \ \ \ \ \ .
\ee
By taking the limit $\epsilon \rightarrow 0$, we obtain \pref{ann1}.
Below, we shall use the corresponding relation for the
spin current to derive boundary conditions on the spin field $\mh$.

The form of the gauge invariant part of the edge action,
${\cal L}^{gi}_{ed}(\Delta^{a}\phi)$,
is not determined by the anomaly
structure. In fact, not even the number of modes is fixed, and in the
above discussion we implicitly assumed a minimal field content on the
edge (this might well change in real systems with flat edge potentials
where edge reconstruction is expected). We shall take the following
Lagrangians,
\be{wen}
{\cal L}_{ed}^{gi} &=& \frac l {4\pi}\left( \Delta_{x}\phi\Delta_0\phi
                    - v(\Delta_x\phi)^2 \right)
        +\langle j^{a}\rangle \Delta_{a}\phi
   \ \ \ \ \ \ \ \ \ \ \ \ \ \ \ \ \ (fp) \label{wen1} \\
{\cal L}_{ed}^{gi } &=& \frac 1 {4\pi}
              \left(l_{\alpha\beta}\Delta_{x}\phi_{\alpha}\Delta_0\phi_\beta
        - v_{\alpha\beta}\Delta_{x}\phi_{\alpha}\Delta_x\phi_\beta \right)
          +\langle j^{a}_{\alpha}\rangle \Delta_{a}\phi_{\alpha}
                       \   \ (pp) \ , \label{wen2}
\ee
where the quadratic parts are identical to the ones proposed by Wen and
others \cite{wen1}, but where we also added explicit terms corresponding to
constant (\ie space and time independent)  ground state charge- and
current densities, for which we shall use the notation
$\rho_{ed} = \langle j^{0}\rangle$ and $j_{ed} = \langle
j^{x}\rangle$ \etc. Note that $\rho_{ed}$
and $j_{ed}$ are parameters in the model that will
depend on the microscopic details of the edge, as discussed below.
Also note that the terms containing $\langle j^{a}\rangle$
do  not contribute to the equation of motion for the field $\phi$,
and in particular that  the mean field ground state for any edge
potential is simply $\phi = 0$.
The velocity $v$, in the (fp) case, and the velocity matrix
$\bf v$, in the (pp) case, are also parameters of the model.
We will use a sign convention where $v>0$.
In the (pp) case, the velocity matrix $\bf v$
can in general be diagonalized simultaneously with $\bf l$ \cite{wen1}.
In our case it is physically quite reasonable to assume that
these two matrices are simply proportional to each other,
${\bf v} = v{\bf l}$, meaning that the edge velocity
of a charged excitation does not depend on its spin.
Although not necessary, we shall for simplicity
restrict our discussion to this special case.
Note that the spin on the edge is given by the boundary value of the
bulk field, so there is no independent spin degree of freedom on the
edge.

The  edge charge and edge current for the (fp) case can now be obtained
from \pref{j1} and \pref{wen1},
\be{fpcurr}
j^{0} &=& \rho_{ed} + \frac l {2\pi} \Delta_{x}\phi
                           \ \ \ \ \ \ \ \ (fp)  \label{fpcurr1} \\
j^{x} &=& j_{ed} - v \frac l {2\pi} \Delta_{x}\phi \label{fpcurr2}
  \ \ \ \ \ ,
\ee
and for the (pp) case \pref{j2} and \pref{wen2} give
\be{ppcurr}
j^{0}_{\alpha} &=& \rho_{ed,\alpha} +  \frac 1 {\pi}
l_{\alpha\beta}\Delta_{x}\phi_{\beta} \ \ \ \ \ \ \  \ \ \ (pp)
                                                 \label{ppcurr1} \\
        j^{x}_{\alpha} &=& j_{ed,\alpha}  -
  v\frac 1 {\pi} l_{\alpha\beta}\Delta_{x}\phi_{\beta} \label{ppcurr2}
  \ \ \ \ \ .
\ee
Note that although the edge theories defined by \pref{ed} and
\pref{wen1} - \pref{wen2} look identical to the spinless case when
expressed in the
covariant derivatives \pref{covdir} and \pref{ppcovdir}, there are in
fact many spin dependent terms, both via $\tilde a$ and via  the
Zeeman term included in $A^{0}$.

We end this section with the following observation.
Since the edge is along the $x$-axis, it is natural to fix the axial
gauge $\tilde a_{y}=0$. For this choice, the bulk charge per unit
length in the $x$-direction due to spin texturing
can be expressed as,
\be{topch}
\rho^{top}(x) = \int_0^{\infty} dy\; (\rho - \rhobar)
            = -\frac{l}{2\pi} \int_0^{\infty} dy\; {\tilde b}
            = -\frac{l}{2\pi} \at_x (x,0) \ \ \ \ \ \  (fp) \ .
\ee
If the topological charge is accumulated along the edge (a concrete
example is given in the next section), it is natural to define the {\em total}
edge charge density by
\be{totedch}
j^0_{tot} \equiv \rho^{top}(x) + j^0 = \rho_{ed} + \frac{l}{2\pi} \p_x \phi
\ \ \ \ (fp) \ \
\ee
(with the gauge choice $A_0 (y=0) = 0$).
In particular, an $x$-independent $\phi$ simply corresponds to a redistribution
of charge in the edge region.
The corresponding expressions for the partially polarized case are
\be{4}
j^0_{tot} &\equiv& \rho^{top}(x)+ j^0_1
 = \rho_{ed,1} + \frac{1}{\pi} l_{1\beta} \p_x \phi_{\beta}
\ \ \ \ (pp) \\
j^0_{tot,s} &\equiv& \rho^{top}_s(x) + j^0_2
 = \rho_{ed,2} + \frac{1}{\pi} l_{2\beta}  \p_x \phi_{\beta}
\ \ \ \ (pp),
\ee
where $\rho^{top}_s(x)$ is the ``spin charge density'' defined by
$\int dy\,(\rho P - \rhobar\Pbar)$.

\vskip 4mm
\begin{center}
{\bf 4. EQUATIONS OF MOTION AND BOUNDARY CONDITIONS}
\end{center}
\renewcommand{\theequation}{4.\arabic{equation}}
\setcounter{equation}{0}
\vskip2mm
In the following we shall treat the fully and partially polarized
cases in parallel. In the partially polarized case, the edge current
$j^{a}$ without subscript will always refer to the spin current $j^{a}_{2}$,
and similarly, $\rho_{ed}$ refers to $\rho_{ed,2}$ and
 $J^{tot}_{\mu}$  to $J^{tot,2}_{\mu}$.

Variation of the total action \pref{stot} with respect to the field $\phi$
(using \pref{ed}, \pref{wen1} and \pref{wen2}) gives
the following equations of motion,
\be{edeom}
\partial_{x} (\Delta_{0} - v\Delta_{x})\phi
                                &=& 0  \ \ \ \ \  (fp) \\
l_{\alpha\beta}\partial_{x} (\Delta_{0} - v\Delta_{x})\phi_{\beta}
                                  &=& 0  \ \ \ \ \  (pp)\ \ \ .  \nonumber
\ee
Variation with respect to the bulk field $\mh$, taking the constraint
$\mh^2 = 1$ properly into account,  gives two
contributions. First the usual ferromagnetic spin wave equation,
\be{bueom}
\partial_{0} \mh - \kappa \mh\times\nabla^2\mh - \mu_e{\vec B}\times\mh = 0
 \ \ \ \ \ ,
\ee
but also an edge part $\sim \delta (y)$ which is not a dynamical
equation, but provides a boundary condition for the spin field $\mh$.
An equivalent and physically more instructive way to obtain this
boundary condition is to require the spin current to be conserved at the
edge:

The total spin current is obtained from the full action $S^{tot}$
using  Noether's theorem and the global symmetry transformation
\pref{symm1},
\be{spincurr}
 \vec J^{tot}_{\mu} = \half \mh J^{tot}_{\mu}
                    -\frac {\kappa\rhobar\Pbar}{2} \mh\times\partial_{\mu}\mh
                                         (1-\delta_{\mu 0})
                                         \theta(y) \ \ \ \ \ ,
        \ee
where the current  $J^{tot}_{\mu}$ is given by \pref{jtot}.
(To derive \pref{spincurr} it is convenient to take variations with
respect to the spinor $\chi_{\mh}$ and use the relation \pref{natilde}.)

As expected, the spin current has a convective part, corresponding to
the  spin of the moving charges, and a ferromagnetic part,
corresponding to spin waves for fixed charge current. The convective
part, being proportional to the total current, has both a bulk
and an edge contribution, while the ferromagnetic component
only exists in the bulk.
In writing \pref{stot} we assumed that the only explicit breaking of the
O(3) spin rotation symmetry comes from the Zeeman term. In particular,
we ignore possible spin-orbit effects, both at the boundary and
in the bulk. The divergence of the spin current is thus given only by
the Zeeman term in $A^0$
(recall $J_0^{tot}A^0 = J_0^{tot}(-\half \mu_e\vec B\cdot\mh
+ \delta A^0)$) and reads
\be{scurrcons}
\p^{\mu}{\vec J}_{\mu}^{tot} = \frac{ J_0^{tot}} {2} \mu_e{\vec B}\times{\mh}
\ \ \ \ \  .
\ee
The boundary condition now follows by using the same arguments as for the
gauge currents (or equivalently by
inserting \pref{spincurr} into \pref{scurrcons} and identifing the terms
$\sim \delta(y)$).  The expression corresponding to
\pref{ann} becomes
\be{bc1}
\partial_{a}(j^{a}\half\mh) +\vec J_{y} = \half j^{0} \mu_{e}
\vec B\times \mh \ \ \ \ \ .
\ee
We now see that the anomaly cancellation conditions \pref{ann}
imply that the $\mh$-component of this equation
is identically satisfied, and the
equations for the remaining two components give the boundary
condition
\be{fullbc}
j^{a} \partial_{a} \mh - \rhobar\Pbar\kappa (\mh \times
\partial_{y}\mh )  - \mu_e j^0 {\vec B}\times{\mh} = 0
  \ \ \ \ \ \ \ \
 \ee
that relates the  $\mh$ field and its derivatives to
the chiral edge current. Had it not been for the anomaly cancellation
that eliminated one of the equations in \pref{bc1}, the equations of
motion would in general only have trivial solutions since there are
only two degrees of freedom in the $\mh$-field.
The boundary condition \pref{fullbc} is one of the main results of this paper.

\vskip 4mm
\begin{center}
{\bf 5. EDGE MODELS AND EDGE MODES }
\end{center}
 \renewcommand{\theequation}{5.\arabic{equation}}
\setcounter{equation}{0}
\vskip2mm \noindent

The boundary condition \pref{fullbc} involves the edge charge and
current densities, and when studying small fluctuations, these can simply
be put equal to their ground state values $\rho_{ed}$ and $j_{ed}$. To
relate these parameters to the edge potential, we shall use a simple
model, where the edge is defined as  a region of size $\Delta y$ where
there is an electric field in the $y$-direction. To be able to
compare our results with those of Karlhede \etal \cite{karlhede3},
we shall follow their construction and divide the external
electrostatic potential into one piece corresponding to
a background charge $A^{0}_{bg}(y)$, which completely cancels the charge of the
electrons in the ground state, and an additional piece $A^{0}$ that
can either tend to contract or expand the edge, according to the
direction of the corresponding electric field. Note that for the
``ideal edge'' where $A^{0} = 0$, there is no net electric field
and thus no drift velocity and no edge drift current in the ground
state. For the non-ideal edge, there is an electric field,
$E_{y} = \partial_{y}A^{0}$  and a corresponding drift velocity
$v_{d} = E_{y}/B_{\perp}$ in the ground state.
If $v_{d}$ is constant over the edge, corresponding to a
constant electric field, the edge charge and edge currents in the
ground state are related by $j_{ed} = v_{d}\rho_{ed}$.
Note that $v_{d}$, which depends on the details of the edge,
can have any sign and is not to be confused with the chiral
velocity $v$, which characterizes the excitations and is always positive.

If we now further assume that the
electron density in the edge region does not differ much from the
bulk value $\rhobar$
(it must of course go to zero at the very edge), we
have 
\be{edI}
\rho_{ed}&=& \Delta y \rhobar  \\
j_{ed} &=&  v_{d} \rho_{ed} =
   \frac {\Delta A^{0} \rhobar} {B_{\perp}} = -\frac \nu {2\pi}\Delta A^{0}
   \ \ \ \ \ ,
\nonumber
\ee
where $\Delta A^{0} = A^{0}(\Delta y) - A^{0}(0)$
is the total potential difference over the edge, and $\nu$ is the
filling fraction.

Note that in this simple model, the edge current is related to the (at
at least in principle) measurable quantity $\Delta A^{0}$. On the
other hand, there is presumbaly no unambiguous way to define the width
$\Delta y$, so $\rho_{ed}$ will remain a free parameter in the model.

In order to study the dynamics of the low energy spin and edge
excitations, we fix the gauge by expressing the  spinor $\chi_{\mh}$ 
in terms of
a small complex number, $z$,
\be{chiz}
\chi_{\mh} =   \left( \begin{array}{c}
        {z} \\ {1-\half|z|^{2}}
\end{array} \right)
\ee
and keep only terms up to ${\cal O}(z^{2})$. The  sigma model fields
are then given by
\be{mhat}
\mh &=& \left[ (z+\zb), i(z-\zb), 2|z|^2 - 1  \right] \nonumber \\
\left(\nabla \mh  \right)^2 
    &=& 4 \left(\nabla\zb \right) \left(\nabla z \right) \\
\at_{\mu} &=& \frac{i}{2} \left(\zb\p_{\mu}z - z\p_{\mu} \zb
\right) \ \ \ \ \ .  \nonumber
\ee
Extracting the $\sigma$-model kinetic energy $\sim\tilde a^0$ and the
Zeeman term included in $A^0$ and $A^0_2$ from \pref{neff1a} and
\pref{neff2},
and using the expressions \pref{mhat}, we get
\be{zL}
{\cal L} = \rhobar\Pbar\; \left[ i\zb \p_0 z + \kappa \zb \nabla^2 z
                     + \mu_e B \left( \half - \zb z \right)
                           \right]
        \ \ \ \ \ ,                                 
\ee
where $\Pbar = 1$ and $B=\left|{\vec B}\right|$ for the fully polarized case.
Substituting the ground state values for the edge charge and current
density into (\ref{fullbc}) gives a boundary condition for $z$,
\be{7}
\left( \rho_{ed}i\p_0 + j_{ed}i\p_x - \rho_{ed}\mu_e B \right) z +
      \kappa\Pbar \rhobar\p_y z = 0
\ \ \ \ .
\ee
(Here we might worry that the relevant edge charge would be
$\langle j^0_{tot}\rangle$, from \pref{totedch} rather
than $\rho_{ed}$, but for the ansatz
\pref{chiz}, the difference is $\sim |A|^2 k$, \ie
of higher order in the small fluctuation amplitude $A$.)

There will be several classes of low-lying modes: bulk
spin waves, edge density waves (which we shall refer to as edge
magnetoplasmons - EMP) and edge spin waves (ESW).

\bigskip
\noindent {\em Bulk spin waves} \\
As expected, there are  bulk spin wave  solutions of eq.\pref{bueom} of the form
\be{bulks}
z &=& e^{i(k x - \omega t)  }\cos (k_y y + \alpha) \\
\phi &=& 0 \ \ \ \ \ \ \ . \nonumber
\ee
The  dispersion relation,
\be{bsw}
\omega = \mu_e B + \kappa (k^2 + k_y^2)\ \ \ ,
\ee
follows immediately from \pref{zL},
and the phase shift $\alpha$ is determined from the boundary condition \pref{7}
\be{phases}
\rho_{ed} \left(\omega    -  \mu_e B \right)  -j_{ed} k \cos\alpha -
 \kappa\Pbar \rhobar k_y \sin\alpha = 0
\ \ \ \ .
\ee
We now turn to the more interesting case of the edge modes.

\bigskip
\noindent {\em Edge magnetoplasmons} \\
These are the well known chiral edge modes discussed by \eg Wen
\cite{wen1}. In our model they appear as solutions where $z=0$, so only
the $\phi$ field is excited.
In the fully polarized case, there is one such mode
with a linear dispersion $\omega = -vk$ where $k<0$. For a model
with the general
form \pref{wen2},
there
are in general two EMP eigenmodes with different velocities $v_{\alpha}$.
However, as already mentioned, in the special case
${\bf v} = v {\bf l}$
they both have the same velocity $v$.

By analyzing the boundary condition \pref{7} it is not hard to see that,
at least in our approximation, there is no solution that simultaneously
excites both the $\phi$ and the $z$ fields, \ie there is no
mixing between the EMP and the ESW.

\bigskip
\noindent {\em Edge spin waves}   \\
In order to study the dispersion of edge spin waves, we now make
the {\em ansatz}
\be{ansatz}
   z &=& A e^{i(kx-\omega t)} e^{-\lambda y} \\
\phi &=& 0\ \ \ \ \ \ \ \ , \nonumber
\ee
with $\lambda > 0$. (\ref{7}) gives
\be{r1}
\rho_{ed}\left( \omega  - \mu_e B \right)  - j_{ed}k - \rhobar\Pbar\kappa\lambda =0
    \ \ \ \ \ .                             \label{r11}
\ee
The equation of motion of $\zb$ corresponding to (\ref{zL}) gives
\be{r2}
\omega = \mu_e B + \kappa k^2  -\kappa \lambda^2 \ \ \ \ \ .
\ee
The dispersion relation is  found by eliminating the parameter $\lambda$
from (\ref{r11})  and (\ref{r2}),
\be{sdr}
\lambda &=& \frac{\rhobar\Pbar}{2\rho_{ed}}
         \left[ -1 + \sqrt{1
         + \left(\frac {2\rho_{ed}} {\rhobar\Pbar} \right)^2\left(
           - \frac{j_{ed}}{\kappa\rho_{ed}}k + k^{2}
          \right)}  \right]   \\
  &=&  - \frac {j_{ed}} {\rhobar\Pbar\kappa} k
    + \frac {\rho_{ed}} {\rhobar\Pbar}
        \left(1-\left( \frac{j_{ed}}{\kappa\rhobar\Pbar} \right)^2\right)k^2
    + {\cal O}(k^{3}) \ \ \ \ \ . \nonumber
\ee
In the generic case where $j_{ed}\neq 0$, we can substitute the model
dependent expression for $j_{ed}$ from \pref{edI} to get
\be{lambda}
\lambda = \frac {\nu \Delta A^{0} } {2\pi\kappa\Pbar\rhobar} k   + {\cal O}(k^{2})
\ \ \ \ \ ,
\ee
and the corresponding dispersion relation
\be{eswdisp}
\omega = \mu_{e} B
       + \kappa \left( 1 - \left( \nu\frac{\Delta A^0}
                         {2\pi\kappa\Pbar\rhobar}\right)^2 \right) k^2
                                                 \ \ \ \ \ ,
\ee
is that of an ESW with a spin stiffness differing from the bulk value $\kappa$.
Note that since $\lambda$ must be positive, the sign of $k$ has to be
the same as that of $\Delta A^{0}$. In particular this means that
for a potential that tends to smoothen the edge, which with our
sign conventions corresponds to $\Delta A^{0} < 0$, we have $k < 0$,
\ie the ESWs propagate in the same direction as the EMPs. This
qualitative result was also found by Karlhede \etal
for the special case of a fully polarized $\nu =1$ state
\cite{karlhede3}. Also,
and perhaps more surprising, if we take $\Pbar = \nu =1$
together with the $\kappa$ value pertinent to Coulomb interaction,
\pref{lambda} exactly coincides with the result of
\cite{karlhede3}.

For the special case of  $\Delta  A^{0} = 0$, corresponding to zero drift
current in the ground state, we get
\be{lambda2}
\lambda = \frac {\rho_{ed}} {\Pbar\rhobar} k^{2}
\ee
and the corresponding dispersion relation
\be{eswdisp2}
\omega = \mu_{e} B + \kappa k^2 \ \ \ \ \ .
\ee
This mode is not chiral, but  exists for both signs of $k$,
and the spin stiffness is not renormalized; corrections
to the bulk spin wave dispersion are ${\cal O}(k^4)$. Both these
qualitative features were found in the $\nu = 1$ case in \cite{karlhede3}.
Since in this case $\lambda$ depends explicitly on
$\rho_{ed}$, we can  not  relate the actual value of $\lambda$
to that in \cite{karlhede3}. It is interesting to note, however,
that in our approach the physical condition for having a  chiral ESW
mode is rather transparent: The convective part of the
spin current \pref{spincurr} has a chiral edge component as soon
as there is a charge drift current in the ground state. Via the
boundary condition \pref{fullbc} this also gives chirality to the
ESWs. When there is no current in the ground state, the ESW is
non-chiral.

Here we should mention that our results differ from those of a
previous sigma model analysis of the fully polarized case by
Milovanovi\'c \cite{milo1}, who \eg finds
a term $\sim k$ in the dispersion relation for the ESW. (The differences
can be traced back to our requirement that the gauge anomaly related
to the Berry gauge field $\tilde a_{\mu}$ should cancel between bulk
and edge.)

We end this section with a discussion of a possible generalization of
the edge Lagrangian.
A basic assumption in constructing the edge theories was that
$\mh$  is continuous at the edge, so that the $\mh$-field in the
boundary theory (that enters ${\cal L}_{ed}$ via $\tilde a_\mu$)
is just the  boundary value of the bulk $\mh$ field, \ie the
spin on the boundary does not constitute an
independent degree of freedom.
Even with this assumption, however, the action $S^{tot}$ in
 \pref{stot} is not the most general one. We  already mentioned
the possibility of extra chiral modes, corresponding to a
(non-textured) reconstruction of the edge, but there is also the
possibility of having an extra $\sigma$-model term in the
Hamiltonian corresponding
to an edge spin stiffness,
\be{edgesigma}
{\cal H}_{\sigma,ed} = \int dx \,
  \frac {\kappa'} 4 j^0 (\partial_{x}\mh(x,0))^{2}
  \ \ \ \ \ . 
\ee
Adding this piece to the previous expression \pref{jtot} for the spin
current and proceeding as before, we find that
the boundary conditions \pref{fullbc} are modified into
\be{8}
j^{a} \partial_{a} \mh
    - \Pbar\rhobar\kappa (\mh \times \partial_{y}\mh )
    - j^0 \kappa' (\mh \times \partial_{x}^2\mh )
    - j^0 \mu_e \vec B \times \mh = 0
  \ \ \ \ \ .
\ee
This gives a correction to the term $\sim k^2$ in the expression
for $\lambda$, \pref{sdr}; in the special case $\kappa' = \kappa$,
the $k^2$ term is cancelled. In the generic case where $j_{ed}\neq 0$, both the
penetration depth, $\lambda^{-1}$, and the ESW dispersion relation,
are unchanged to leading order, while
for the non-chiral case, $j_{ed}=0$, $\lambda$ depends sensitively
on $\kappa'$.
This further emphasizes our previous comment that the non-chiral case can
only be qualitatively described within our model.


\vskip 35mm
\begin{center}
{\bf 6. SUMMARY AND OUTLOOK }
\end{center}
\renewcommand{\theequation}{5.\arabic{equation}}
\setcounter{equation}{0}
\vskip2mm
Since the derivation of the combined effective theory for the bulk and edge
is rather technical, we want
to summarize the main steps in order to highlight the rather simple and
compelling logic.

First the bulk theories are derived
from the CSGL-Lagrangians by integrating out the hard density and polarization
modes (for details, se Appendix A). The resulting effective actions
depend on the background
electromagnetic field and a single dynamical spin field, $\mh$. We make a
derivative expansion and neglect terms with more than two derivatives of
$\mh$.  Zeeman terms are included as spin dependent scalar potentials.
For the partially polarized case there are two gauge invariances, one related
to electromagnetism and another related to the choice of spin basis. In the
case of full polarization, these gauge invariances coincide.

When restricting the bulk
theory to a half-plane, there are gauge non-invariant terms in the action,
so we add a non-gauge invariant edge action such that the combined
theory is invariant under both gauge transformations. This implies that
both the charge current and the $\mh$-component of the spin current are
conserved. The gauge invariant part of the edge action is not fixed by
invariance arguments, and we choose a minimal model,
\pref{wen1} and \pref{wen2}, which
describes the edge as a chiral Luttinger liquid.

Assuming that the global O(3) spin symmetry is violated only by the Zeeman
term, we derive a boundary condition for the spin field $\mh$.
Together with the equations of motion for
$\mh$ in the bulk and $\phi$ on the boundary, this boundary condition
defines the combined bulk - edge theory. It is important
that in this model the spins on the boundary do not constitute
any independent degrees of freedom.

Finally, we did a small fluctuation analysis of the full model and found a
spectrum in agreement with the $\nu = 1$ Hartree-Fock results
in Ref.\cite{karlhede3}. It is encouraging that our results, which were
derived in a very general context
and based mainly on symmetry considerations, give the same results as
a specific, much more controlled calculation, both  qualitatively
(number of modes and their chirality) and to some
extent also quantitatively (the exact expression for the penetration depth
in the chiral case). This makes it plausible
that our effective action really
does describe the charge and spin-dynamics of  quantum Hall edges,
at non-integer filling fractions and/or partial polarization.

A natural extension of the present work
would be to study textured edges
\cite{karlhede1,oaknin1,leinaas1,karlhede2,franco1}.
To do this one would have
to keep the potential terms that we neglected since they included higher
derivatives. Such terms would  affect the boundary condition, but in
principle it is straightforward to write it down together with the
corresponding Hamiltonian. Since, for certain choices of the parameters,
the ESW equation  \pref{eswdisp} can become soft for large $k$, it is not
inconceivable that texturing the edge will lower the energy. By expanding
the action around such a static solution, one could then go on to study
the interplay between the expected gapless spin waves and the
EMP modes.

A  potentially interesting application of the theory developed in this
paper, would be to study the interplay between EMP modes and spin
waves in a narrow strip. By expanding our Lagrangian to higher order
in small fluctuations, one
generates interaction vertices between the EMP:s and the spin
waves and one could then use perturbation theory to calculate an
effective spin-mediated interaction between the EMP:s at the two
edges.

\vskip 3mm\noi {\bf Acknowledgement}: We thank Anders Karlhede and
Jon Magne Leinaas for helpful discussions and comments on the manuscript.
T.H.H. acknowledges financial support by the Swedish Natural 
Science Research Council.

\vskip 4mm
\begin{center}
{\bf APPENDIX A: FROM CS LAGRANGIANS TO EFFECTIVE $\sigma$-MODELS }
\end{center}
\renewcommand{\theequation}{A.\arabic{equation}}
\setcounter{equation}{0}
\vskip2mm

In this Appendix we shall outline the derivation of the effective Lagrangians
\pref{neff1a} and \pref{neff2}. We consider three different cases. The first,
idealized case of spinless (sl) electrons is relevant for systems
with a very large
Zeeman gap and is included for reference. We then turn to the
fully polarized (fp) and then to the partially polarized (pp) states.
To make the presentation self-contained, we reproduce some relevant
definitions and
equations from section 2.

We shall describe all three systems with the following Lagrangian of the
CSGL type,
\be{slag}
{\cal L} = {\cal L}_\phi + {\cal L}_{CS}(a^{\mu}_{\alpha})
- {\cal H}_\sigma .
\ee
In the partially polarized case, $\phi$ is a complex bosonic spinor describing
the electrons, in the fully polarized and spinless models it is just a
single complex field,
and $a^{\mu}_{\alpha}$ are auxiliary CS fields \cite{hansson1}.
The sigma model term
${\cal H}_\sigma$ is absent in the spinless case.
In all cases we have
\be{lagp}
{\cal L}_\phi = \phidag iD_0 \phi -\frac 1 {2m_{e}} |\vec D\phi|^2
- V(\rho)
 \ \ \ \ \ ,
\ee
where $m_{e}$ is the electron mass, and $\rho = \phidag\phi$ is the
density.
The exact form of the potential $V(\rho)$ varies for the different cases,
but is of
no importance for the following discussion.
For the spinless  and the fully polarized cases the covariant
derivative and the
CS Lagrangian are given by
\be{cs1}
iD^\mu &=& i\partial^\mu + a^\mu_1 + A^\mu \ \ \ \ \ \ \ \ (sl) \label{cov1} \\
iD^\mu &=& i\partial^\mu + a^\mu_1 + A^\mu + \tilde a^{\mu} \ \ (fp)
   \label{cov1b} \\
{\cal L}_{CS} &=& - \frac l {4\pi}
\epsilon_{\mu\nu\sigma}a_{1}^\mu\partial^\nu a_{1}^\sigma \equiv
-\frac l {4\pi}  a_{1}da_{1}  \ \ \ \ \ .
\ee
Please recall that the three-vector notation is only
for notational convenience; the metric
is Euclidian, \ie $D^{\mu} = D_{\mu}$, and all signs are written explicitly.
$l^{-1}$ is an odd integer, and the topological vector potential
$\tilde a$
is defined by
\be{atilde}
\tilde a^\mu = \chidag_{\mh} i\partial^\mu \chi_{\mh}
 \ \ \ \ \ ,
\ee
with the spinor $\chi_{\mh} $  defined by
\be{chi}
\hat m^i = \chidag_{\mh} \sigma^i \chi_{\mh} \ \ \ \ \ ,
\ee
up to a gauge transformation.
As discussed in \cite{hansson1}, the potential $\tilde a^\mu$
represents the Berry phases picked up
by the spins on the electrons, and the gauge freedom corresponds
to the freedom
of picking the local spin basis. The spinless case differs from
the fully polarized
only in that the gauge field $\tilde a^\mu$
is absent.
The vector field $\hat m$, which is present in addition to the
field $\phi$ and the CS fields
$a_\alpha^\mu$,
describes the direction of the spin polarization.
The corresponding expressions in the partially polarized case are
\be{cs2}
iD^\mu &=& i\partial^\mu + a_1^\mu + a_2^\mu\sz + A^\mu  -\half
(\mh\times\partial^\mu \mh)\cdot\vec\sigma \ \ \ \ \  \label{cov2} \\
{\cal L}_{CS} &=&  - \frac 1 {2\pi} \ki \alpha \beta
   \epsilon_{\mu\nu\sigma}a_\alpha^\mu\partial^\nu a_\beta^\sigma
   \equiv  - \frac 1 {2\pi} \ki \alpha \beta a_{\alpha}da_{\beta}
 \ \ \ \ \ ,
\ee
where the elements of the symmetric matrix
${\bf l}^{-1}$ are integers, with diagonal elements being
both either even or odd.

The corresponding
sigma model terms for the fully and partially
polarized cases are both given by
\be{sigma}
{\cal H}_\sigma  = \frac {V_0} 2 (\partial _i \vec S)^2
 + \mu_{e}  \vec S \cdot \vec B \ \ \ \ \ ,
\ee
where $-\mu_{e} = -eg/(2m)$ is the effective
magnetic moment of the electron, $V_0$ is an interaction parameter,
and the spin density $\vec S$, is related to the unit vector field
$\mh$, by  $\vec S = (1/2) \,\mh \, \phidag
\vec \sigma \cdot \mh \,\phi$.

The pertinent symmetries of the Lagrangian \pref{slag} are given by
\be{symm}
\chi \rightarrow e^{i\alpha(x)}\chi \ \ , \ \ \ a_{1}^{\mu}
&\rightarrow& a_{1}^{\mu} + \partial^{\mu}\alpha(x)  \label{symm1} \\
\chi \rightarrow e^{i\beta(x)\vec \sigma \cdot \mh (x)}\chi \ \ , \ \ \
a_{2}^{\mu} &\rightarrow&
a_{2}^{\mu} + \partial^{\mu}\beta(x) \label{symm2} \\
\chi \rightarrow  e^{\frac i 2 \vec k\cdot\vec\sigma}\chi  \ \ , \ \ \ \mh
&\rightarrow&
e^{i\vec k\cdot \vec L}\mh  \ \ \  \ \ , \label{symm3}
\ee
where the $3 \times 3$ matrix $\vec L$ is the  angular
momentum in the vector representation.

The gauge symmetry \pref{symm1} connected to $a_1$ is related
to electric charge conservation and is present in
all three cases. The gauge symmetry \pref{symm2}  connected
to $a_2$ is relevant  only to the
partially polarized case and is related to having another conserved
charge corresponding to the local polarization density, as discussed
in some detail
in \cite{hansson1}. There it is also stressed that while the
charge corresponding to $a_1$ is the usual electric charge,
and the correponding gauge-symmetry is fundamental,
the second charge is only conserved
under certain physical assumptions, namely neglecting spin-flip
transitions. In this paper we simply assume that
the partially polarized states can be described by the above theory with two
gauge symmetries.

For theories with spin and in the absence of a  Zeeman interaction,
there is also a global symmetry \pref{symm3}, corresponding to O(3) spin
rotations.

The mean field ground state in the spinless case
is given by \cite{zhang1}
\be{gnstfp}
\rhobar = \langle \rho\rangle =  \frac l {2\pi} b_{1}
  =  - \frac l {2\pi}  B_\perp \ \ \ \ \ ,
\ee
with filling fraction $\nu = l$.
For the fully polarized case, the same condition holds, and the O(3) symmetry
is spontaneously broken so that the $\hat m$ field points in a fixed direction
$\hat m_0$. The corresponding Goldstone bosons are ferromagnetic spin waves,
and in the presence of a symmetry breaking
Zeeman term they get a Zeeman gap $\mu_e | \vec B |$.

For the partially polarized states the mean field ground state is given by
\be{gnst}
\bar\rho &=& -\ki 1 1 B_\perp /\pi  \nonumber \\
\bar P &=& \langle \hat n\cdot \hat m \rangle  =\cos\bar\alpha =
{\ki 1 2 }/{\ki 1 1 } \ \ \ \ \ ,
\ee
with filling fraction $\nu= 2{\ki 1 1} $
and polarization $\bar P $.

We now assume that the $g$-factor is small enough for the
spin waves to be  considered as low-energy excitations, while there
are  large (cyclotron) gaps to density fluctuations. In the partially
polarized case there are two density modes corresponding to
independent fluctuations of spin up and spin down electrons, or
equivalently, one density and one polarization mode. To get an
effective low energy theory, the hard modes are integrated out.
Technically one proceeds by making a phase - density
decomposition of the fields ($\phi = \sqrt\rho e^{i\theta}$ in the (fp)
case and  a similar expression for the (pp) case given by \pref{decom}
in Appendix B), fixing unitary gauge, substituting in \pref{lagp}
and neglecting all derivatives of the density and polarization to get
\cite{hansson1}
\be{efflag}
{\cal L}_{\phi} &=& \rho(a^0 + A^{0} + \tilde a^0 )
 - V(\rho) -\frac \rho {2m_{e}} (\vec a_1 + \vec A +
 \atvidv)^2 \ \ \ \ (fp)  \label{efflag1} \\
{\cal L}_\phi &=& \rho[\aeo + A^{0} + \ca\ato + \ca\tilde a^0 ]
   - V(\rup,\rdown)  \ \ \ \ \ \ \ \ \   (pp)   \nonumber   \\
    &-& \frac \rho {2m_{e}} \{[\aev + \vec A + \ca (\atv + \atvidv) ]^2
+ \sin^{2}\alpha (\atv + \atvidv )^2 \} \ \ \ , \label{efflag2}
\ee
where both $A^\mu$ and $\tilde a^\mu$ are considered as external
fields. It is now easy to show that the modes corresponding to the
gauge fields $a^{\mu}_{\alpha}$ have gaps proportional to the
cyclotron energy and can be integrated out.
A simple way to obtain the resulting Lagrangian is
to first calculate the currents corresponding to the two gauge fields
in the mean field approximation and then integrate
these expressions\cite{zhang1}. Starting with the spinless case, we
get,
\be{bcurr1}
J_\mu \equiv \frac {\delta S} {\delta A^\mu}
      = \frac {\delta S_\phi} {\delta a_1^\mu}
          = - \frac {\delta S_{cs} } {\delta a_1^\mu}
        = \frac l {2\pi} \epsilon_{\mu\nu\sigma}\partial^{\nu} a^{\sigma}_{1}
          = - \frac l {2\pi} \epsilon_{\mu\nu\sigma}\partial^{\nu} A^{\sigma}
                  \ \ \ ,                                  
\ee
where $S$ is the action corresponding to ${\cal L}$ \etc, and where we used
the equation of motion $\delta S/\delta a_{1}^{\mu}=0$ and the mean
field  condition $a_1^\mu = - A^\mu$. We can now  integrate with respect to
$A^\mu$ to get the result for spinless electrons,
\be{csact}
{\cal L}^{sl}_{eff} = - \frac l {4\pi} AdA - V(\rhobar) \ \ \ \ \  .
\ee

Noting that \pref{efflag1} depends on $\tilde a$ only through the
combination $A + {\tilde a}$, the result for the fully
polarized case now simply follows by
making  the substitution $A \rightarrow A + \tilde a$ and adding the sigma
model term,
\be{eff1}
{\cal L}^{fp}_{eff} = -\frac l {4\pi}  (A + \tilde a )d(A + \tilde
a)- V(\rho) -{\cal H}_{\sigma} \ \ \ \ \  .
\ee
Since we ignored density fluctuations in deriving this effective
Lagrangian, it cannot describe vortices,
and is thus appropriate only
for energies well below the threshold for creation of
Laughlin type quasiparticles. It does, however, describe
spin-texture and thus skyrmions.
(It is also not difficult to extend the
effective Lagrangian as to include vortex currents, as discussed in
Appendix B.)

Next we rewrite \pref{eff1} in a way that is suitable for constructing
the edge theory. Note that the Zeeman term in ${\cal H}_{\sigma}$ is directly
proportional to the density, while the spin stiffness term is proportional to
the density squared. In the approximation we use, the density fluctuations are
proportional to $\tilde b$ and thus to second derivatives of the $\mh$ field.
Keeping terms only to quadratic order in derivatives of $\mh$, we can thus make
the replacement $\rho \rightarrow \rhobar$ in the spin stiffness term and in
the potential term (remember that $V(\rho) = V(\rhobar) + O(\tilde b^2)$,
but we have to keep $\rho$ in the Zeeman term. Since a term  proportional to
the density is  nothing but an electrostatic potential, we can incorporate the
Zeeman term by shifting the  potential $A^0$ as
\be{anoll}
A^0 \rightarrow A^0 - \half \mu_{e}\vec B\cdot \mh \ \ \ \ \  .
\ee
Since gauge invariance will be a major guiding principle for
constructing the edge theories, this way
of including the Zeeman term is very advantageous, and throughout
the paper we
assume the shift \pref{anoll} of $A^0$ if not stated otherwise.
We shall also neglect the constant term $V(\rhobar)$, so, to summarize,
we have the following effective bulk Lagrangian,
\be{eff1a}
{\cal L}^{fp}_{eff} = -\frac l {4\pi}  (A + \tilde a )d(A + \tilde
a) - \frac {\kappa\rhobar} 4 (\vec\nabla\mh)^2  \ \ \ \ \  .
\ee

It is straightforward to make a similar construction
in the partially polarized case by noting that
the gauge fields in \pref{efflag2} only occur in the combinations
$A + a_{1}$ and $\tilde a + a_{2}$.
Here we have two conserved currents
corresponding to
the two external  potentials $A^\mu$ and
$\tilde a^\mu$. Using the convenient notation $A_1 = A$
and $ A_2 = \tilde a$ and performing
the same manipulations as in \pref{bcurr1}, gives, {\em mutatis
mutandis},
\be{bcurr2}
J_\alpha^\mu  = - \frac 1 \pi \ki\alpha\beta
              \epsilon^{\mu\nu\sigma} d_{\nu}
              A_{\sigma}^\beta \ \ \ \ \  ,                   
\ee
which can be integrated and combined with the $\sigma$-model action to give
\be{eff2}
{\cal L}^{pp}_{eff} =  -\frac 1 {2\pi} \ki \alpha\beta A_\alpha d A_\beta
- \frac {\kappa\rhobar\Pbar} 4 (\vec\nabla\mh)^2
 \ \ \ \ \ .
\ee
In this case the Zeeman term is included by a shift in the potential $A_2^0$.
Note that in both cases, the  topological vector potential $\tilde
a^\mu$ is simply  related to the ferromagnetic kinetic term by
\be{ferro}
{\cal L}_{kin} &=& \rhobar\Pbar  \tilde a^{0}  \ \ \ \ \ ,
\ee
(see, \eg \cite{stone1}),
so using the mean field solutions \pref{gnstfp} and \pref{gnst},
the term $\sim \tilde a^0 B$ in \pref{eff1a} and \pref{eff2}
provides the correct kinetic term for the sigma model field $\hat m$.
To get  \pref{ferro}, we again substituted $\rho \rightarrow\rhobar$, but
it is important to note that the higher derivative term
$\sim \tilde ad\tilde a$ is very important when going beyond linear response.
It  is a so-called Hopf term, which is normalized  such that skyrmions,
which are topological soliton solutions to \pref{eff1a} and \pref{eff2}, get
the correct (fractional) statistics\cite{hansson2,hansson1}.

\vskip 4mm
\begin{center}
{\bf APPENDIX B: DUAL CS THEORY FOR PARTIALLY POLARIZED STATES }
\end{center}
\renewcommand{\theequation}{B.\arabic{equation}}
\setcounter{equation}{0}
\vskip2mm
Here we present the dual version of the CSGL model for partially
polarized QH states and show how it can be used to give an alternative
derivation of the low-energy effective bulk theory discussed in
section 2. The advantage of this approach is that vortex currents
are included in a very straightforward way.
In this duality transformation, which is a direct generalization
of the one for spinless electrons (as described in \eg \cite{zhang2}),
one goes from the original
CSGL model describing electrons in an external field, to a description
where the quasiparticle excitations (vortices) are considered to be
the fundamental particles.

Consider the ${\cal L}_{\phi}+ {\cal L}_{CS}$ part of eq.(\ref{slag})
(${\cal H}_{\sigma}$ does not
enter the duality transformation and is
omitted for simplicity), and decompose the matter field as follows (for
notation, see \cite{hansson1}),
\be{decom}
\phi = \sqrt\rho e^{i\vartheta}e^{i\frac \alpha 2  \vec\sigma\cdot
\hat {e}_1 (\theta) } \chi_{\mh} \ \ \ \ \ .
\ee
As in the text, we neglect gradient terms in the density $\rho$ and
polarization $\alpha$, but we decompose the phase angles $\vartheta$
and $\theta$ into regular parts that are absorbed in the gauge fields $a_1$
and $a_2$, and singular parts $\eta_1$ and $\eta_2$ that describe
vortices,
\be{decomp}
\vartheta = \vartheta_{reg} + \eta_{1} + \eta_{2} \\
\theta = \theta_{reg} - 2\eta_{2} \nonumber \ \ \ \ \ .
\ee
The singular phase angles are defined
by $\oint d\vec r\, \vec\nabla \eta_{i} = 2\pi n$ for an infinitesimal
contour enclosing the singularity.
Doing this, it is a matter of algebra to derive \cite{hansson1},
\be{cslag}
{\cal L} &=& \rho \left[ a_{01} + \p_0 \eta_1
         +\cos\alpha \left(a_{02} +\at_0 + \p_0 \eta_2  \right)
         + A_0 \right] \nonumber \\
        &-& \frac{\rho}{2m} \left[ \left( a_{i1} + \p_i \eta_1
         +\cos\alpha \left(a_{i2} +\at_i +\p_i \eta_2  \right)
         + A_i \right)^2 \right. \nonumber \\
        &+& \left. \sin^{2}\alpha\left(a_{i2} +\at_i +\p_i\eta_2\right)^2
         \right]  - V \nonumber \\
        &-& \frac{1}{2\pi} l_{\alpha\beta} \epsilon_{\mu\nu\sigma}
           a_{\mu}^{\alpha}\p_{\nu} a_{\sigma}^{\beta} \ \ \ \ \ .
\ee
Next, the space components of the CS fields are  split
into their longitudinal and transverse parts,
\be{gaugedec}
a_{i}^{\alpha} = \partial_{i}\theta_{\alpha}
       + \varepsilon_{ij}\partial_{j} \phi_{\alpha}
              \ \ \ \ \ .
\ee
Integrating out $a_{0}^{\alpha}$ gives the CS constraints
\be{ACS}
\rho &=& \frac{1}{\pi} l_{1\beta}b_{\beta}  \label{ACS2} \\
\rho\cos\alpha &=& \frac{1}{\pi} l_{2\beta}b_{\beta} \ \ \ \ \ . \label{ACS3}
\ee
The quadratic terms in the Lagrangian are linearized by introducing
Hubbard-Stratonovic
fields $J^{\mu}_{\alpha}$, such that
\be{LL}
{\cal L} &=& \rho \left[ A_0 + \p_0 \eta_1
         +\cos\alpha \left(\at_0 +\p_0 \eta_2  \right)
         \right]
          + \frac{m}{2\rho} \left(J_1^2 + J_2^2  \right)  \\
         &+& J_{1i} \left[ A_i + \epsilon_{ij}\p_j\phi_1 + \p_i\theta_1
          + \p_i \eta_1
          + \cos\alpha\left( \at_i + \epsilon_{ij}\p_j\phi_2
          + \p_i\theta_2 +\p_i\eta_2 \right)\right] \nonumber \\
         &+& J_{2i} \sin\alpha \left[\at_i + \epsilon_{ij}\p_j\phi_2
          + \p_i\theta_2 + \p_i\eta_2  \right] - V \nonumber \\
         &-& \frac{1}{\pi} l_{\alpha\beta}
             \phi_{\alpha} \nabla^2\p_0 \theta_{\beta}
          \nonumber  \ \ \ \ \ .
\ee
Integrating out the phase fields $\theta_{\alpha}$ gives the
constraints
$\partial_{\mu}J^{\mu}_{C} =\partial_{\mu}J^{\mu}_{S} =0 $
where the two conserved
currents are
\be{acurr}
J^{\mu}_C &=& \left(\rho, {\vec J}_1  \right)
          \equiv \epsilon_{\mu\nu\sigma} \p_{\nu} B_{\sigma}^{(1)}
                                                 \\
J^{\mu}_S &=& \left(\rho\cos\alpha,{\vec J}_1 \cos\alpha
           + {\vec J}_2 \sin\alpha  \right)
          \equiv \epsilon_{\mu\nu\sigma} \p_{\nu} B_{\sigma}^{(2)}
                  \ \ \ \ \ .
\ee
Here, we have explicitly solved the constraints by introducing
{\em dual Chern-Simons fields} $B_{\mu}^{\alpha}$.

\noindent
Using this definition along with (\ref{gaugedec}) - (\ref{ACS3}),
the fields $\phi_{\alpha}$ can be  expressed in
terms of the dual CS fields as
\be{pa}
\phi_{\alpha} = 
- \frac{\pi}{\nabla^2}
               l^{-1}_{\alpha\beta} \epsilon^{ij} \p_i B_j^{\beta}
                             \ \ \ \ \  .
\ee
Thus eliminating $\phi_{\alpha}$ from $\cal L$ and introducing the
vortex currents
\be{vcurr}
{\hat j}^{\mu}_{\alpha} &\equiv& \frac{1}{2\pi} \epsilon^{\mu\nu\sigma}
                  \p_{\nu} \p_{\sigma}\eta_{\alpha}  \ \ \ \ \ ,
\ee
(note that this expression is not zero because of the singularity in
$\eta_{\alpha}$),
one finds the final expression for the complete dual bulk action,
\be{dba}
{\cal L} &=& \frac{\pi}{2} ~ l^{-1}_{\alpha\beta} ~
     \epsilon^{\mu\nu\sigma}B_{\mu}^{\alpha} \p_{\nu} B_{\sigma}^{\beta}
         + \epsilon^{\mu\nu\sigma}
           A_{\mu}^{\alpha} \p_{\nu} B_{\sigma\alpha} \nonumber
         + 2\pi  B_{\mu}^{\alpha} {\hat j}^{\mu}_{\alpha}  \\
        &+& F^2  - V(\rho)     \label{duaL}
\ee
where $F^2$ represents terms which are higher order in derivatives
and will be neglected from now on.

\bigskip
\noindent
The effective action is obtained by integrating out the dual CS fields
in (\ref{duaL}). To this end, we rewrite the gauge part of (\ref{duaL})
as
\be{deffa}
{\cal L} &=& \frac{\pi}{2} l^{-1}_{\alpha\beta} ~
     \epsilon^{\mu\nu\sigma}B_{\mu}^{\alpha} \p_{\nu} B_{\sigma}^{\beta}
          + B_{\mu}^{\alpha}{\cal J}^{\mu}_{\alpha} \\
         &\equiv& -\half B_{\mu}^{\alpha}
                         \left( G_{\alpha\beta}^{\mu\nu}  \right)^{-1}
                         B_{\nu}^{\beta}
          + B_{\mu}^{\alpha}{\cal J}^{\mu}_{\alpha} \nonumber
\ee
where
\be{ducurr}
{\cal J}_{\mu}^{\alpha}\equiv 2\pi {\hat j}_{\mu}^{\alpha}
        +  \epsilon^{\mu\nu\sigma} \p_{\nu} A_{\sigma}^{\alpha}  \label{dualJ}
\ee
and
\be{fields}
\left( G_{\alpha\beta}^{\mu\nu}  \right)^{-1}
  = \pi l^{-1}_{\alpha\beta}\epsilon^{\mu\nu\sigma} \p_{\sigma} \ \ \ \ \  .
\ee
A convenient gauge to choose for the external field is
$\p^{\mu}A_{\mu}^{(1)}=0$.
In this gauge, the propagator is
\be{prop}
G_{\alpha\beta}^{\mu\nu} = -\frac{1}{\pi} l_{\alpha\beta}
              \epsilon^{\mu\nu\sigma} \frac{\p_{\sigma}}{\p^2}
                          \ \ \ \ \ . \label{dualG}
\ee
Upon integrating out the fields $B_{\mu}^{\alpha}$, the action reduces to
\be{duact1}
{\cal L}_{eff} &=& \half {\cal J}_{\mu}^{\alpha}
                         G_{\alpha\beta}^{\mu\nu}
                         {\cal J}_{\nu}^{\beta} \ \ \ \ \  .
\ee
Inserting for ${\cal J}_{\mu}^{\alpha}$ and $ G_{\alpha\beta}^{\mu\nu}$
from (\ref{dualJ}) and (\ref{dualG})
and neglecting terms $\sim ({\hat j})^2$
finally leads to the effective bulk action, in the notation used in
section 2,
\be{duact2}
{\cal L}_{eff} = -\frac{1}{2\pi} l_{\alpha\beta} A_{\alpha} d A_{\beta}
               - l_{\alpha\beta} \left(
                 A^{\alpha}{\hat j}^{\beta}
               + A^{\beta}{\hat j}^{\alpha} \right) \ \ \ \ \ .
\ee
We see that in the absence of vortex currents, this reduces to the
gauge part of the action (\ref{neff2}) found from the original CSGL model.

\end{document}